\documentclass[10pt,conference]{IEEEtran}
\IEEEoverridecommandlockouts
\usepackage{cite}
\usepackage{amsmath,amssymb,amsfonts}
\usepackage{algorithm}
\usepackage[noend]{algpseudocode}
\usepackage{algorithmicx}
\usepackage{graphicx}
\usepackage{textcomp}
\usepackage{xcolor}
\usepackage{booktabs}
\usepackage{epstopdf}
\usepackage[T1]{fontenc}
\usepackage{subfigure}
\usepackage{comment}
\def\BibTeX{{\rm B\kern-.05em{\sc i\kern-.025em b}\kern-.08em
    T\kern-.1667em\lower.7ex\hbox{E}\kern-.125emX}}
\begin{document}

\title{End-Edge Coordinated Joint Encoding and Neural Enhancement for Low-Light Video Analytics\\
}
\author{\IEEEauthorblockN{Yuanyi~He\IEEEauthorrefmark{1},~Peng~Yang\IEEEauthorrefmark{1},~Tian~Qin\IEEEauthorrefmark{1}, and~Ning~Zhang\IEEEauthorrefmark{2}}
  \IEEEauthorblockA{\IEEEauthorrefmark{1}School of Electronic Information and Communications, Huazhong University of Science and Technology, Wuhan, China \\
   \IEEEauthorrefmark{2}Department of Electrical and Computer Engineering, University of Windsor, Windsor, ON, Canada\\
    Email: \IEEEauthorrefmark{1}\{yuanyi$\_$he, yangpeng, qin$\_$tian\}@hust.edu.cn, \IEEEauthorrefmark{2}ning.zhang@uwindsor.ca}
   }
\begin{comment}
\author{\IEEEauthorblockN{1\textsuperscript{st} Given Name Surname}
\IEEEauthorblockA{\textit{dept. name of organization (of Aff.)} \\
\textit{name of organization (of Aff.)}\\
City, Country \\
email address or ORCID}
\and
\IEEEauthorblockN{2\textsuperscript{nd} Given Name Surname}
\IEEEauthorblockA{\textit{dept. name of organization (of Aff.)} \\
\textit{name of organization (of Aff.)}\\
City, Country \\
email address or ORCID}
\and
\IEEEauthorblockN{3\textsuperscript{rd} Given Name Surname}
\IEEEauthorblockA{\textit{dept. name of organization (of Aff.)} \\
\textit{name of organization (of Aff.)}\\
City, Country \\
email address or ORCID}
\and
\IEEEauthorblockN{4\textsuperscript{th} Given Name Surname}
\IEEEauthorblockA{\textit{dept. name of organization (of Aff.)} \\
\textit{name of organization (of Aff.)}\\
City, Country \\
email address or ORCID}
\and
\IEEEauthorblockN{5\textsuperscript{th} Given Name Surname}
\IEEEauthorblockA{\textit{dept. name of organization (of Aff.)} \\
\textit{name of organization (of Aff.)}\\
City, Country \\
email address or ORCID}
\and
\IEEEauthorblockN{6\textsuperscript{th} Given Name Surname}
\IEEEauthorblockA{\textit{dept. name of organization (of Aff.)} \\
\textit{name of organization (of Aff.)}\\
City, Country \\
email address or ORCID}
}	
\end{comment}

\maketitle

\begin{abstract}
In this paper, we investigate video analytics in low-light environments, and propose an end-edge coordinated system with joint video encoding and enhancement. It adaptively transmits low-light videos from cameras and performs enhancement and inference tasks at the edge. Firstly, according to our observations, both encoding and enhancement for low-light videos have a significant impact on inference accuracy, which directly influences bandwidth and computation overhead. Secondly, due to the limitation of built-in computation resources, cameras perform encoding and transmitting frames to the edge. The edge executes neural enhancement to process low contrast, detail loss, and color distortion on low-light videos before inference. Finally, an adaptive controller is designed at the edge to select quantization parameters and scales of neural enhancement networks, aiming to improve the inference accuracy and meet the latency requirements. Extensive real-world experiments demonstrate that, the proposed system can achieve a better trade-off between communication and computation resources and optimize the inference accuracy.
\end{abstract}
\begin{comment}
\begin{IEEEkeywords}
component, formatting, style, styling, insert
\end{IEEEkeywords}	
\end{comment}

\section{Introduction}
With the progress of computer vision technologies, many unprecedented applications based on real world vision have emerged in recent years. It holds great potential for 
advancing a broad range of scientific and commercial applications\cite{olatunji2019video}. Visual analysis models have 
been widely used in traffic, monitoring, surveillance, retail, factory\cite{jiawei}, etc.,  in order to provide efficient and safe services and bring economic benefits\cite{du2020server}.
Despite the fact that state-of-the-art deep neural networks (DNNs) can achieve stunning accuracy in video analytics tasks\cite{yangpeng}, the limited computing resources in the camera 
are insufficient to execute expensive native DNNs. 

To optimize the performance of DNNs, it has been proposed that local video analysis tasks should be offloaded to edge nodes, whereby video frames are transmitted to nearby edge nodes 
in pursuit of more favorable computation resources\cite{chengyan}. This scheme allows for efficient distribution and allocation of computation resources between the camera and edge, reducing latency and data 
transmission while facilitating the incorporation of advanced machine learning algorithms into resource-constrained simulations\cite{ananthanarayanan2019video,Chuqin}. 
At the same time, the bandwidth resources available for transmission to the edge are limited.
Most existing works have tweaked the 
video by adjusting the encoding parameters like frame rate, resolution, quantization parameter (QP), or a combination of them. For instance, Jiang \emph{et al.}\cite{jiang2018chameleon} and  
Liu \emph{et al.}\cite{10.1145/3503161.3548033} show that dynamic adjustment of encoding parameters can significantly reduce bandwidth consumption without reducing analytics accuracy, 
while Wang \emph{et al.}\cite{9155524} show that adaptive control of resolution and frame rate is also beneficial to improve transmission efficiency and energy saving. 
In spite of their effectiveness, existing works have failed to take full advantages of edge nodes for enhancement in the face of low-light environments, a common phenomenon appearing at night due to insufficient illumination\cite{Kong}.

Due to the increasing difficulty of performing visual tasks under low-light conditions compared to normal-light scenarios, nighttime autonomous driving and surveillance video analysis are burgeoning research areas that pose significant technical challenges. The quality of video can be severely degraded under low-light conditions, leading to diminished visibility, reduction in detail, and color distortion. These factors pose substantial challenges for visual recognition and analysis tasks. Therefore, high level visual tasks in low-light conditions more difficult than normal-light conditions \cite{wang2022self}.

Although various low-light enhancement algorithms have been proposed, 
they cannot be directly employed in video analytics. Firstly, traditional 
approaches are based on distribution mapping to expands the range of an image, such as histogram equalization \cite{coltuc2006exact}. Another general approach relies on Retinex theory to estimate the corresponding illumination map of an image \cite{7782813}. These two methods bring color distortion hence hardly meet the accuracy improvement requirements. Secondly, methods based on supervised training such as \cite{tao2017llcnn} require paired data, which is intensive to acquire in real-time video analytics. Therefore, it is necessary to utilize unsupervised learning methods\cite{jiang2021enlightengan, guo2020zero}. Thirdly, enhancement can often be excessive for DNN inference without providing proper adjustment and accommodating the content characteristics of video.

In this paper, we explore the effective utilization of low-light enhancement techniques in the context of low-light visual analysis tasks. 
Our investigations entail the enhancement of neural networks implemented with unsupervised learning methods. An end-edge coordinated joint encoding and neural enhancement video analytics system is proposed, aiming to adaptively stream and enhance 
the video captured in low-light environment with proper codec configuration and fine-tuned enhancement model.
The effects of different QP values on bandwidth consumption and inference accuracy of the decoded frames are firstly revealed. Then the computation overhead 
of changing the scale of the enhancement model (i.e., the number of enhancement neural network channels per convolution layer) is performed along with the inference accuracy of the enhancement frames. During handling low-light scenes at runtime, the online adaptive configuration is further customized for accuracy and latency according to video content. 
We design an adaptive configuration selection algorithm, which can efficiently reduce the latency and improve the inference accuracy, thus 
achieving the trade-off between communication and computing resources. In practice, our system can be applied to the areas with low-light conditions, such as parking lots, alleys, or outdoor spaces at night-time.
Our main contributions are summarized as follows.
\begin{itemize}
	\item We conduct experiments with real-world low-light videos to reveal the impact of QP and enhancement models on video analysis accuracy, as well as the communication and computation 
	overhead involved.
	\item We propose a novel optimized low-light video analytics system. Through camera encoding and edge enhancement
	the low-light video is adaptively streamed under communication and computation constraints and the inference accuracy is improved.
	\item We design a heuristic algorithm based on simulated annealing algorithm to solve the configuration selection problem, which has been verified through real-world simulations.
\end{itemize}

The remainder of this paper is as follows. In Section II, we present our motivation and experimental discovery. 
Section III explains our system model and problem formulation, as well as our proposed algorithm. Section IV presents 
evaluation of the system based on realistic videos. We conclude the paper in Section V.

\begin{figure}[t]
	\centering
	\subfigure[]{
		\begin{minipage}[b]{0.22\textwidth}
			\includegraphics[width=1\textwidth]{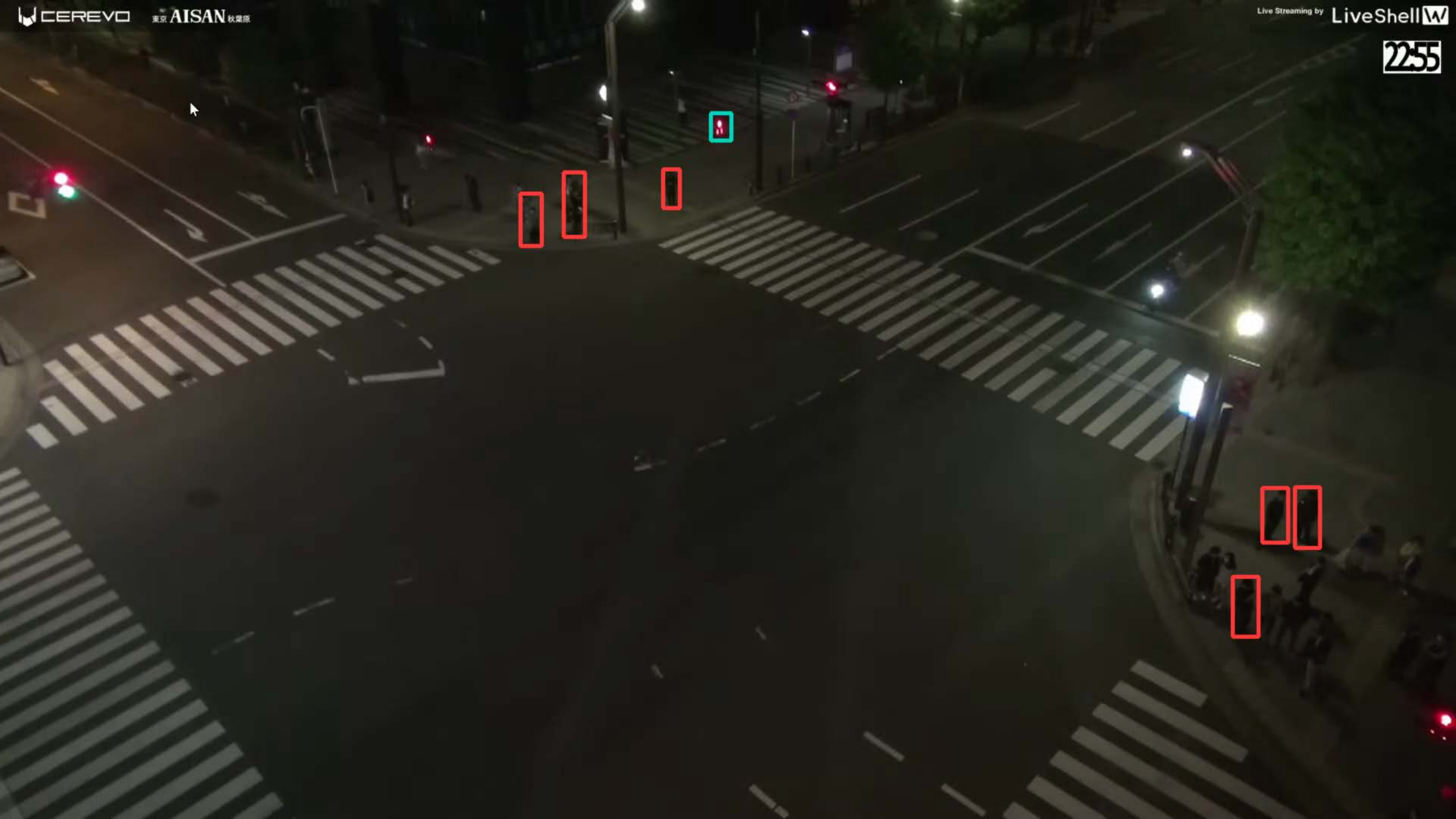} 
		\end{minipage}
		\label{Low-light frame}
	}
    	\subfigure[]{
    		\begin{minipage}[b]{0.22\textwidth}
		 	\includegraphics[width=1\textwidth]{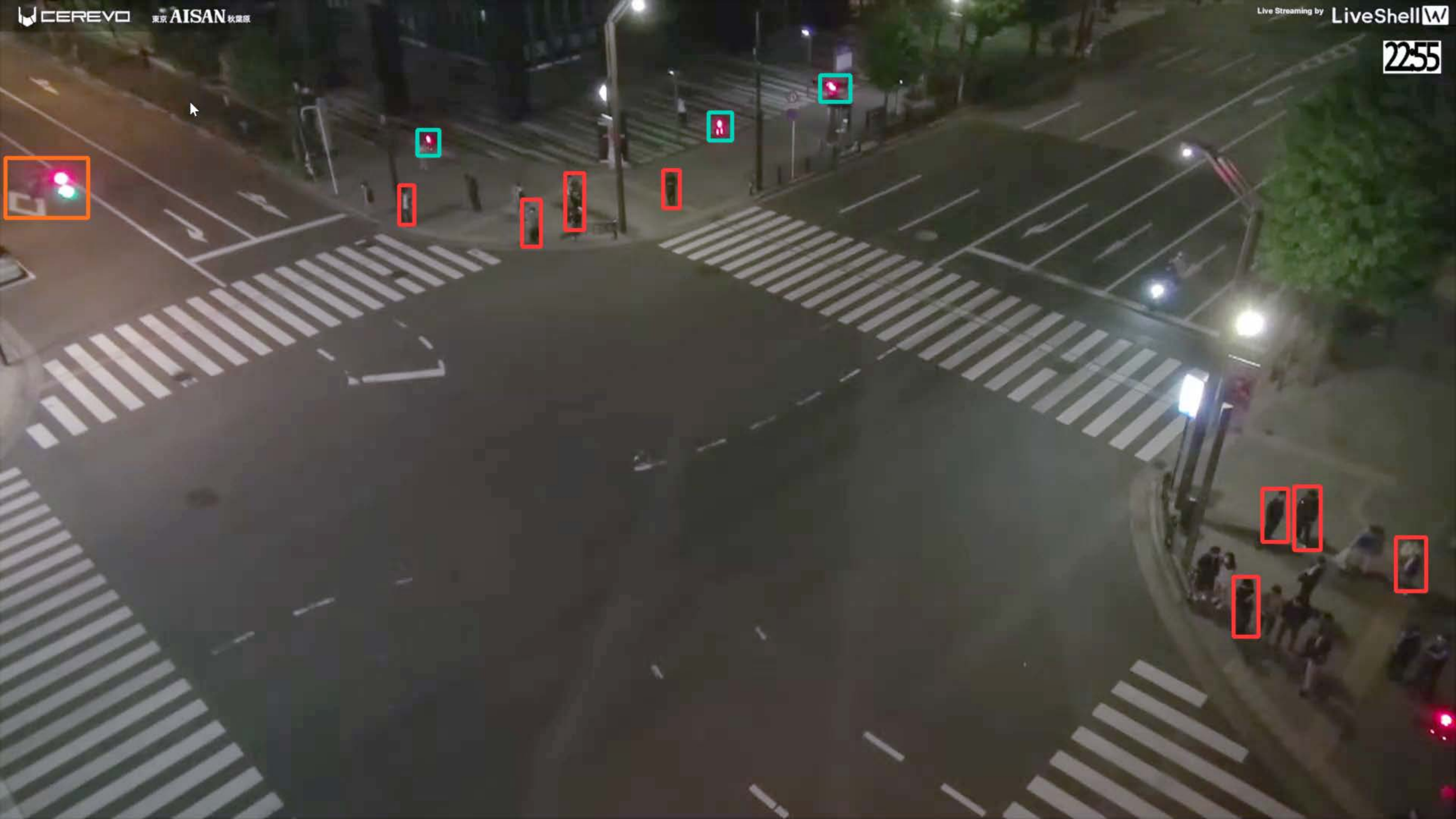}
    		\end{minipage}
		\label{Enhanced frame pi 0.25}
	}
	\subfigure[]{
		\begin{minipage}[b]{0.22\textwidth}
			\includegraphics[width=1\textwidth]{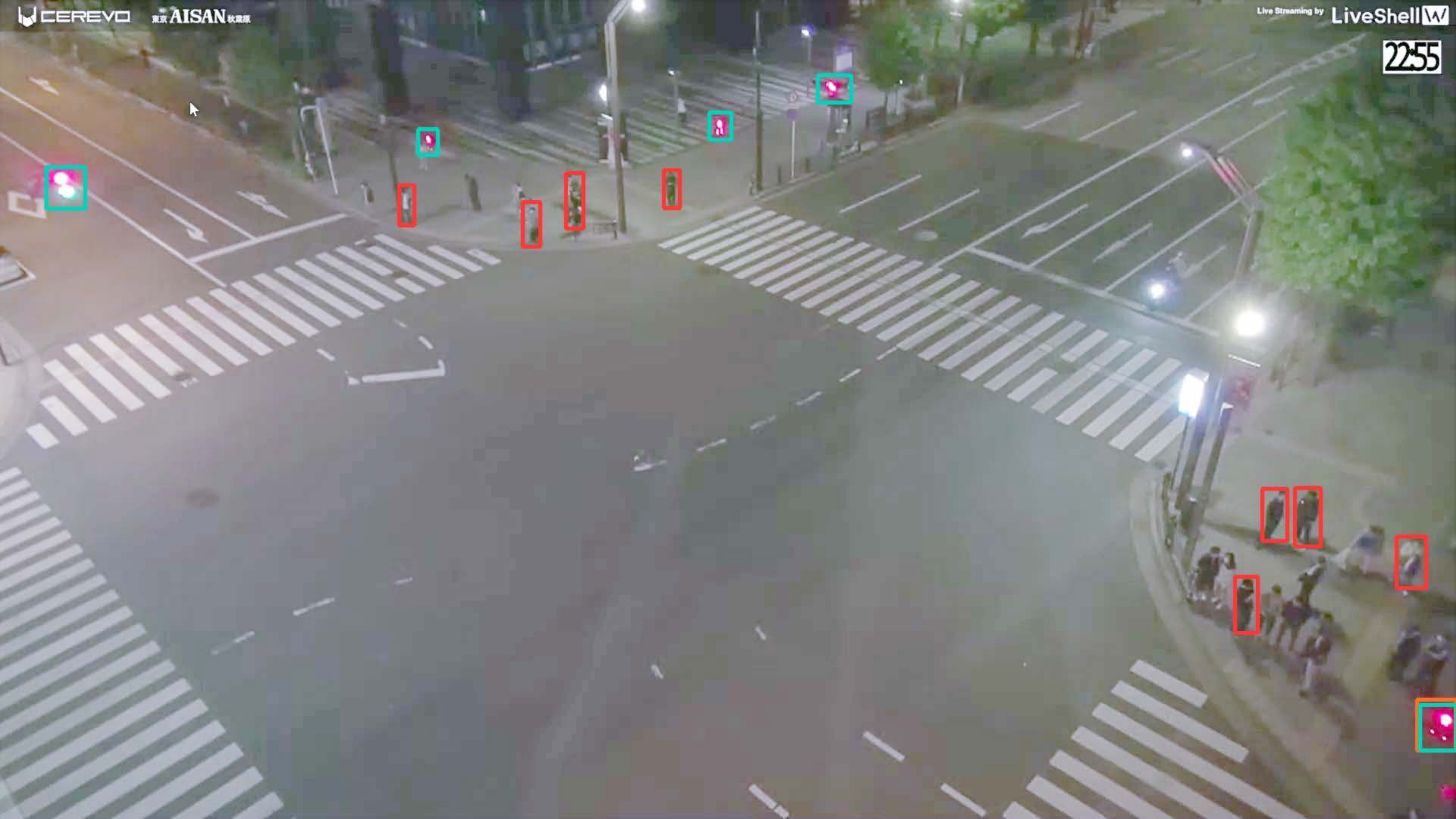} 
		\end{minipage}
		\label{Enhanced frame pi 0.75}
	}
		\subfigure[]{
			\begin{minipage}[b]{0.22\textwidth}
				\includegraphics[width=1\textwidth]{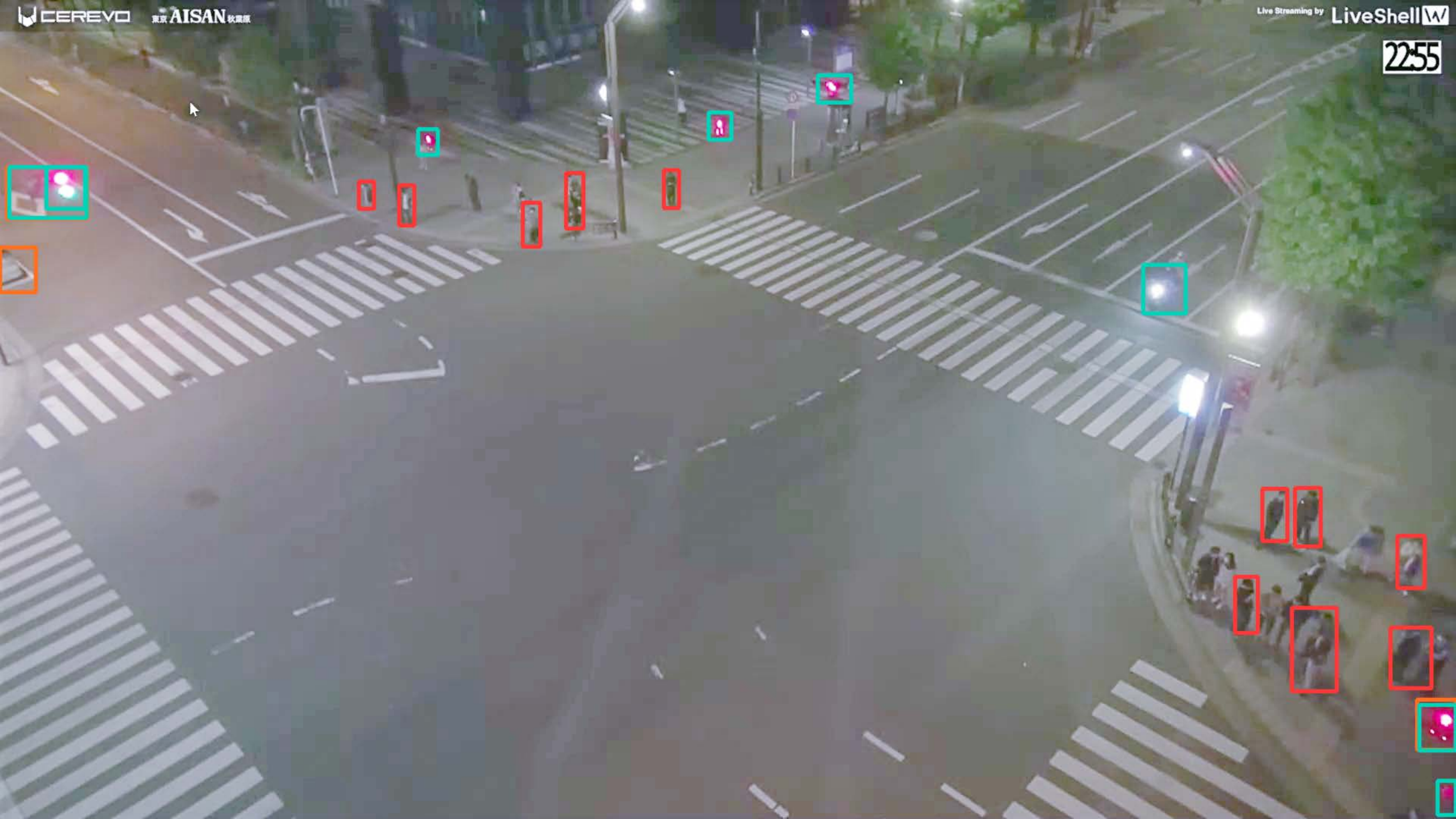}
			\end{minipage}
		\label{Enhanced frame}
		}
	\caption{Comparison of low-light frame and enhanced frame. (a) is low-light frame and (b)-(d) are enhanced frames with the model scale ratio 0.25, 0.75 and 1.}
	\vspace{-3.7mm}
	\label{fig:grid_4figs_1cap_4subcap}
\end{figure}

\section{Motivation}
\subsection{Encoding Low-Light Video}
Most mobile and built-in platforms have equipped with H.264 video codec with standardized implementations. 
As one of efficient encoding pipelines, H.264 supports plenty of adjustable parameters to achieve effective 
lossy compression and encoding\cite{10.1145/3503161.3548033}. One of the most regular used one is QP, 
indicating the video quality. Quantization is performed after Discrete Cosine Transform (DCT), reducing 
the precision of the transform coefficients based on QP\cite{richardson2011h}. Unlike the normal-light videos, videos captured in low-light environment suffer from noise with short exposure as shown in Fig. \ref{Low-light frame}.
Despite camera with higher ISO increasing brightness, it introduces noise as well\cite{chen2018learning}. 
This will not only disturb DNN inference but also enlarge datasize, bringing difficulties for transmission. 
DCT can convert the input residual block from the time domain to the frequency domain. This is because the coefficients of high 
amplitude are usually closely related to noise and they can be diminished by setting the QP to a large value. 
However, this will also filter out the details in the frame. 

\begin{figure}[t]
	\centering
	\subfigure[The effects of QP on accuracy and datasize.]
	{
		\begin{minipage}[b]{.45\linewidth}
			\centering
			\includegraphics[scale=0.228]{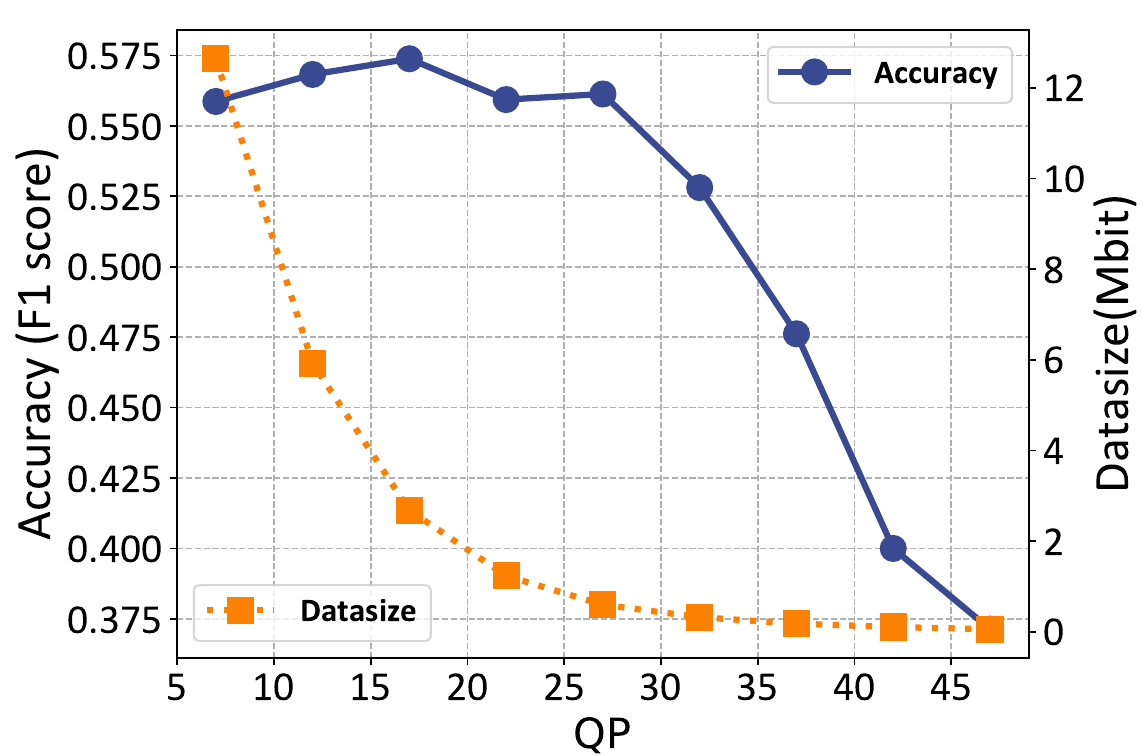}
			\vspace{-4mm}
			\label{The effects of QP on accuracy and datasize}
		\end{minipage}
	}
	\subfigure[The effects of model scale ratio on accuracy and FLOPS.]
	{
		\begin{minipage}[b]{.45\linewidth}
			\centering
			\includegraphics[scale=0.2345]{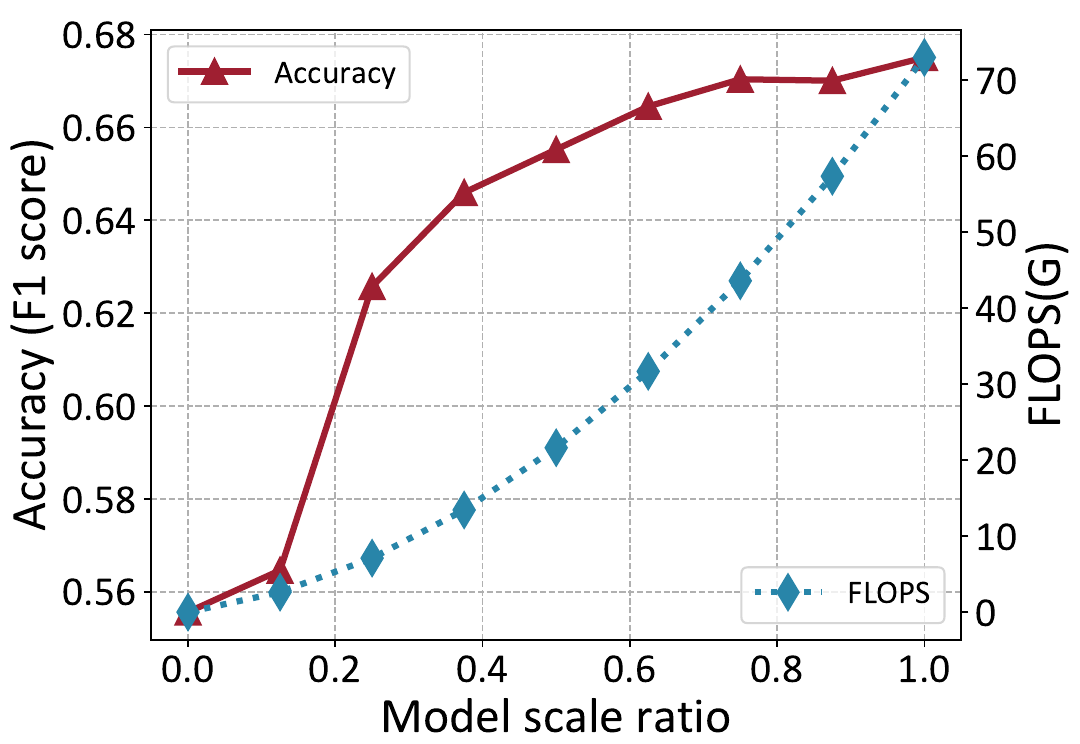}
			\vspace{-4mm}
			\label{The effects of channels on accuracy and FLOPS}
		\end{minipage}
	}
	\caption{Observation on QP and model scale ratio of the enhancement network.}
	\label{Ovservation on QP and enhancement level}
	\vspace{-5mm}
\end{figure}

 We then conduct the widely used object detection model YOLOv5\footnote{
	https://github.com/ultralytics/yolov5. Accessed May 18, 2023.
 }
 on a video clip of a highway surveillance camera from YouTube\footnote{
	https://www.youtube.com/watch?v=slOgQojt8w8. Accessed May 18, 2023.
 }
 with various QP values. The duration of the clip is 1 second. We measure the datasize after encoding with different QP. The evaluation of accuracy is F1 
 score, the harmonic mean of precision and recall of the detected objects between encoded frames and unencoded frames
 \cite{9155524}. As shown in Fig. \ref{The effects of QP 
 on accuracy and datasize}, higher QP can increase low-light video compression ratios (i.e., smaller datasize) while 
 degrading the inference accuracy. This characteristic is similar to normal-light video. In addition, the accuracy 
 decreases when the QP value is less than 17. This is because some noise is not filtered out, thus affecting the 
 inference.
\subsection{Low-light Enhancement}
Enhancement is the counterpart of encoding. The camera can denoise the captured video and reduce the transmission 
datasize by quantization in encoding, while the edge can enhance the decoded video to adjust the brightness, contrast 
and color of the video and reconstruct the details.

Next, we conduct object detection on the decoded frames enhanced by Zero-DCE \cite{guo2020zero}, a deep-curve mapping with a lightweight DNN that is trained to estimate pixel-wise and high-order curves for adjusting the dynamic range of an input image. The model is trained on the dataset \cite{cai2018learning}. 
 To fine-tune the enhancement network, we utilize sub-model training mechanism proposed in \cite{li2023fast}. We compare the detected objects between sub-model enhanced frames and full-model enhanced frames. The frame enhanced by full-model is shown in Fig. \ref{Enhanced frame}. 
 And the enhanced frames enhanced by sub-models with the model scale ratio 0.25 and 0.75 are shown in Fig. \ref{Enhanced frame pi 0.25} and Fig. \ref{Enhanced frame pi 0.75}. We use Floating-point Operations (FLOPs) to evaluate the computation overhead for enhancement.
 We explore the inference accuracy of enhancing frames encoded at a certain QP value and enhanced by sub-model with different number of channels
 (i.e., convolution layer kernels in Zero-DCE network). We use the model scale ratio to represent the sub-model channels compared to the full-model channels. 
As shown in Fig. \ref{The effects of channels on accuracy and FLOPS}, the model scale ratio  
has a significant impact on the accuracy and computation. 
Specifically, as the network channels grows, the accuracy of analytics improves and the improvements become gentle at a large number
of channels. Apart from that, the enhancement model with more number of channels also has a larger FLOPs value, 
which means that the model requires more computation resources.
\begin{comment}
Above all, accuracy improvements can be achieved by encoding with smaller QP values or enhancement with larger number of network channels. 
High quality encoding increases transmission overhead while enhancement with big model introduces additional computation consumption. 
Therefore, it is important to trade off transmission and computation to obtain more accuracy gains.
\end{comment}

\section{System Model and Problem Formulation}

\begin{figure}[t]
	\centerline{\includegraphics[scale=0.26]{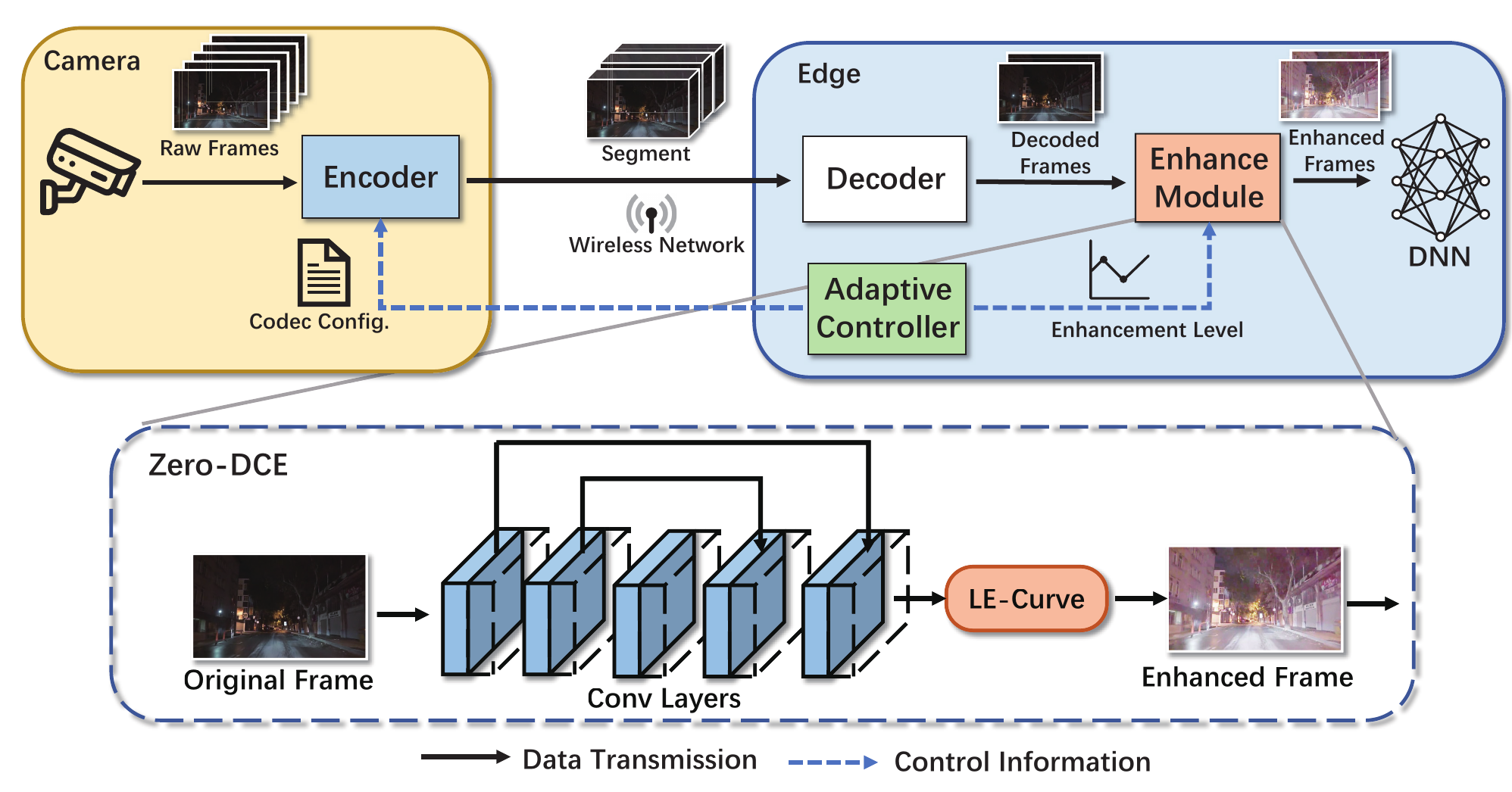}}
	%\vspace{-3.8mm}
	\caption{The system overview.}
	\label{fig_system_model}
\end{figure}
In this section, we first present the system model of low-light environment video analytics, then formulate the joint 
quality level and enhancement level selection problem.

\subsection{System Model}\label{AA}
Considering an end-edge coordinated video analytics scenario that includes one edge server with an access point and multiple stationary surveillance cameras, we design an end-edge coordinated enhancement video analytics system in Fig. \ref{fig_system_model}. The core objective of our system is 
to transmit the encoded video frames from the cameras to the edge for enhancement and detection. In the system workflow, raw frames are temporally 
encoded into several segments with the same duration, and the segments are transmitted through wireless network and 
decoded at the edge. Then low-light enhancement will be first employed to handle the defective frames before 
the inference. Specifically, the built-in hardware on cameras is insufficient, which requires to offload video inference tasks to the edge\cite{10000829}. 
Two main modules are designed at the edge, i.e., the enhancement module and the adaptive controller. The 
enhancement module enhances the low-light frames by neural network approach, here we use Zero-DCE in our model, 
which is a state-of-the-art low-light enhancement network. Obviously, the enhancement network can be changed 
according to the actual situation. 

Formally, the whole video is encoded into $T$ segments for transmission considering the temporal continuity. The 
set $\mathcal{T}=\left\{ 1,\cdots ,t,\cdots ,T \right\} $ denotes the video segment index, where $1\le t\le T$. 
Each video segment contains several fixed frames encoded at the camera side. There are $N$ enhancement 
levels, and the elements of set $\mathcal{K}=\left\{ k_0,k_1,\cdots ,k_n,\cdots ,k_N \right\}$ denote the model scale ratio of 
the enhancement network at corresponding enhancement 
levels, where $0\le n\le N$ and $k_0$ means no enhancement is performed. Higher enhancement levels correspond to 
larger model scale ratio. Also, there are $M$ quality levels, 
corresponding to the quantization parameters. The set $\mathcal{Q}=\left\{ q_1,\cdots ,q_m,\cdots ,q_M \right\}$ 
denotes the quantization parameters at all quality levels, where $1\le m\le M$. Higher quality levels correspond 
to smaller quantization parameters. The adaptive controller selects the appropriate enhancement level and quality level to fully inspire the ability of the system to perform visual tasks in low-light environments.

\subsection{Accuracy and Latency Model}

\begin{figure}[t]
	\centerline{\includegraphics[scale=0.33]{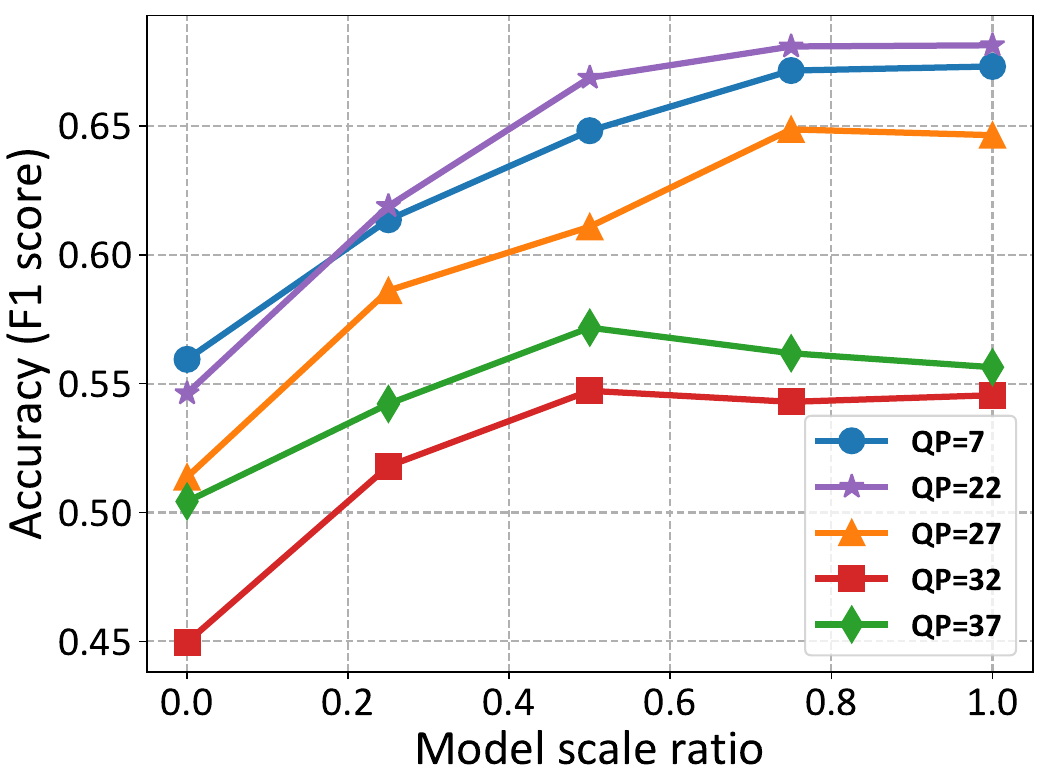}}
	%\vspace{-3.8mm}
	\caption{The accuracy with different QP and model scale ratio.}
	\label{fig_accu_qp_ch}
	%\vspace{-5mm}
\end{figure}

It is observed that enhancement level and quality level have dependent impact on accuracy. Even though frames with higher 
quality level can improve accuracy, high enhancement level does not necessarily result in accuracy gains. As shown in 
Fig. \ref{fig_accu_qp_ch}, when the quantization parameter equals to 22 and 27, the increase of enhancement level makes the accuracy 
improvement. When the quantization parameter equals to 32 and 37, the accuracy decreases with higher enhancement level. The 
configuration $\left(QP=7\right)$ has similar performance to the configuration $\left(QP=22\right)$, while the former takes more 
bandwidth to transmit. Without loss of generality, the accuracy is related both enhancement level and quality level, 
denoted as $A_t\left( k_t,q_t \right)$ representing the accuracy of $t$-th segment.

The system latency comes from many aspects. First, the encoding delay $l_{1}$ and decoding delay $l_{2}$ are related 
to quantization parameters, which can be denoted as $l_1\left( q_t \right)$ and $l_2\left( q_t \right)$ in $t$-th 
segment. When the quantization parameters increases, it results in shorter encoding time. %and less transmission cost but the accuracy 
%drops at the same time. 
The transmission delay $l_3$ is formulated as follows
\begin{equation}
	l_3\left( B_t,q_t \right) =\frac{D\left(q_t \right)}{B_t},\label{eq1}
\end{equation}
where $D\left( q_t \right)$ represents the incremental function of encoded datasize as the quality level grows, and $B_t$ represents 
the maximum transmission bitrate between the camera and the edge under bandwidth constraint in segment $t$. 

The enhancement introduces additional computation overhead. Note that the % including delay and energy consumption
value of FLOPS is constant to enhance a frame with certain resolution. The enhancement computation delay $l_{4}$ can be formulated as bellow
\begin{equation}
	l_4\left( k_t \right) =\frac{f\cdot W\left( k_t \right)}{C},\label{eq2}
\end{equation}
where $W\left( k_t \right)$ denotes the computing workloads of enhancement level $k_t$ at the edge 
in segment $t$ and $C$ represents the computing frequency for enhancing one frame. $f$ is the number of frames in segment $t$. %$\varepsilon \left( \cdot \right)$ 
%is a binary function, where the value remains one when the variable is positive otherwise it remains zero. It is 
%obvious that the function is a switch of computaion delay if the enhancement task is executed.
\begin{comment}
The corresponding energy consumption of the enhancement is shown as follows
\begin{equation}
	E_t=\epsilon fC^2\cdot W\left( k_t \right) \label{eq3}
\end{equation}
where $\epsilon$ is the energy coefficient of the edge depending on the GPU architecture. %, and $P$ represents the 
%operating power consumption at the edge.	
\end{comment} 
Then the overall latency in $t$-th segment can be formulated by
\begin{equation}
	L_t=l_1\left(q_t \right)+l_2\left(q_t \right) +l_3\left( B_t,q_t \right)  + l_{4}\left( k_t  \right)+L_p,\label{eq4}
\end{equation}
 where the DNN inference delay is denoted as $L_{p}$, which is a constant in our system.

\subsection{Problem Formulation}
As for adaptive controller, it is required to select the appropriate enhancement level $k_t$ and quality level $q_t$ for 
$t$-th segment in the situation of variable and limited bandwidth and computation resources to balance the trade-off 
between accuracy and latency. 
Our goal is to maximize the long-term utility function of the accuracy and the latency through the whole video. The 
problem can be expressed as follows
\begin{equation}
	\label{eq5}
	\begin{split}
		\mathcal{P}:\underset{k_t\in \mathcal{K},q_t\in \mathcal{Q}}{\max}\,\,\sum_{t=1}^T&{ \left[A_t\left( q_t,k_t \right) -\lambda L_t\right] }  \\
		\text{s.t.\,\,}C_1:\,\,k_{t}&\in \mathcal{K},\,\,\forall t\in \mathcal{T}, \\
		C_2:\,\, q_t&\in \mathcal{Q},\,\,\forall t\in \mathcal{T}, \\
		C_3:\,\,L_t&\le L_0,\,\,\forall t\in \mathcal{T}.\\
	\end{split}			
\end{equation}
$\lambda \in \left[ 0,1 \right]$ is the weight %$A_r$ denotes the lower bound of the accuracy requirement and 
parameter accounting for the balance between the accuracy and the latency. $E_0$ and $B_0$ are energy and bitrate constraint, respectively. $C_1$ and $C_2$ indicate the 
range of selection for enhancement level and quality level. $C_3$ represents that the latency for transmission and computation
in segment $t$ should not exceed the constraint $L_0$. Problem $\mathcal{P}$ is a mixed integer non-linear programming (MINP), and it is
not feasible to solve it through a brute force approach whose computation complexity is $O\left( \left( MN \right) ^T \right) $. Therefore, we need to
design an lightweight algorithm to solve problem $\mathcal{P}$.
%lower than the accuracy requirement. 

\subsection{Algorithm Design}
\begin{algorithm}[t]
	\caption{SA-based Configuration Selection}
    \label{alg:sa}
    \begin{algorithmic}[1]
        \Require Initial $c=[q,k]$, function $\hat{A_t}, B_t, W, C, f$
        \State Initialize current temperature $T$ and cooling coefficient $\alpha$
		\State Compute utility $U(c\,; \hat{A_t}, B_t, W, C, f)$
        \While{$T > T_{min}$}
            \For{$t=1$ to $limit_T$}
                \State Generate neighboring configuration $c^{\prime}$ of $c$
				\If{$GetLatency(c) \le L_0$}
					\State Compute utility $U(c^{\prime}; \hat{A_t}, B_t, W, C, f)$
					\State $\Delta U = U(c^{\prime}) - U(c)$
					\If{$\Delta U > 0$}
						\State Accept new configuration $c^{\prime}$
					\Else
						\State Generate random number $p \in [0, 1]$
						\If{$p < \exp(-\Delta U / T)$}
							\State Accept new configuration $c^{\prime}$
						\EndIf
					\EndIf
				\EndIf
            \EndFor
            \State $T = T \times \alpha$
        \EndWhile
        \State \Return Best configuration found
    \end{algorithmic}
	\vspace{-1mm}
\end {algorithm}
To efficiently solve the problem $\mathcal{P}$, we propose a SA-based quality level and enhancement level selection algorithm, which serves as a heuristic 
optimization algorithm aiming to search for a sub-optimal configuration for each segment respectively. 
It iteratively searches for better solutions by randomly modifying the current solution and accepts them only if they improve 
or satisfy Metropolis Criterion\cite{metropolis1953equation}. The probability of accepting worse solutions depends on a temperature parameter 
in order to explore the search space more broadly. As the algorithm progresses, the temperature parameter decreases gradually until it reaches 
the termination condition and the decrease speed is related to the cooling coefficient $\alpha$, where $\alpha \in (0, 1)$. The utility function in our algorithm
is denoted as follows
\begin{equation}
	U=\hat{A_t}\left( q_t,k_t \right) -\lambda L_t\left( q_t,k_t,B_t,W,C,f\right),\label{eq6}
\end{equation}
where $\hat{A_t}\left( q_t,k_t \right)$ is the estimated accuracy function. Our proposed algorithm is summarized in
Algorithm 1. Line 3-15 contain an outer loop that terminates when a certain temperature threshold is reached, while 
line 4-14 include an inner loop that determines the number of candidate values generated during each iteration.

\section{Performance Evaluation}
In this section, we evaluate the performance of our system for low-light environment. The experimental settings and results are presented as follows.
\begin{figure*}[ht]
	\centering
	\subfigure[The performance on average accuracy.]
	{
		\begin{minipage}[b]{.3\linewidth}
			\centering
			\includegraphics[scale=0.27]{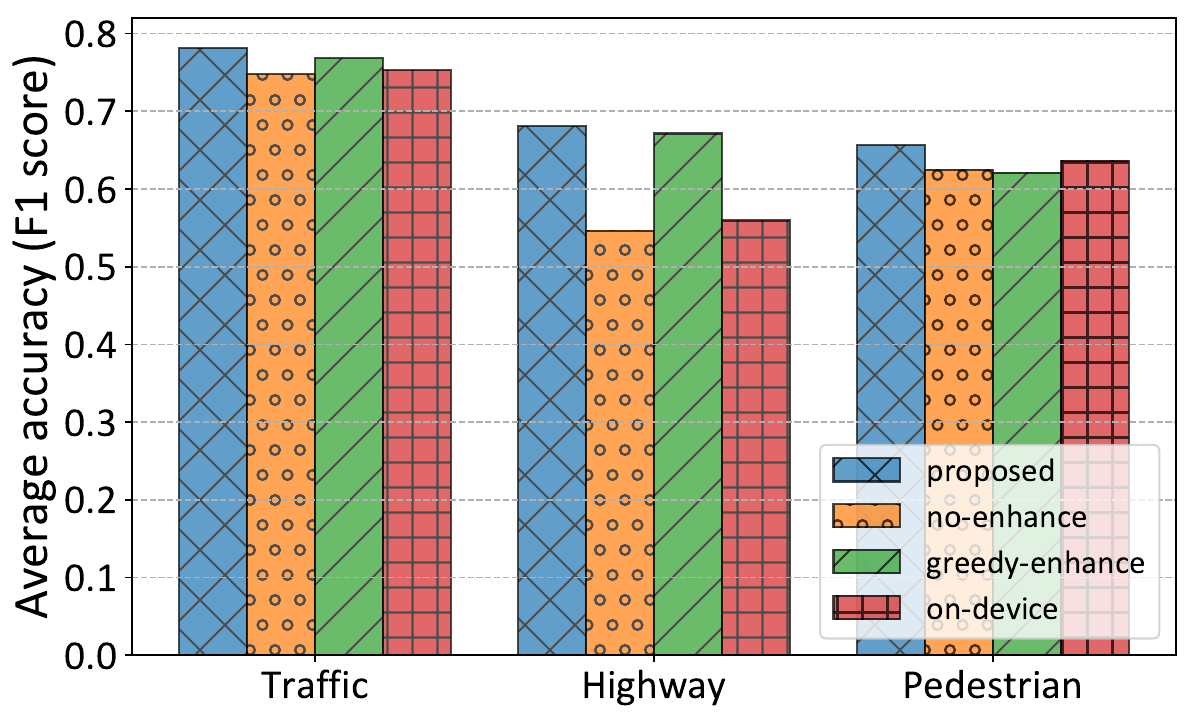}
			\label{accu_method}
			\vspace{-5mm}
		\end{minipage}
	}
	\subfigure[The performance on average latency.]
	{
		 \begin{minipage}[b]{.3\linewidth}
			\centering
			\includegraphics[scale=0.28]{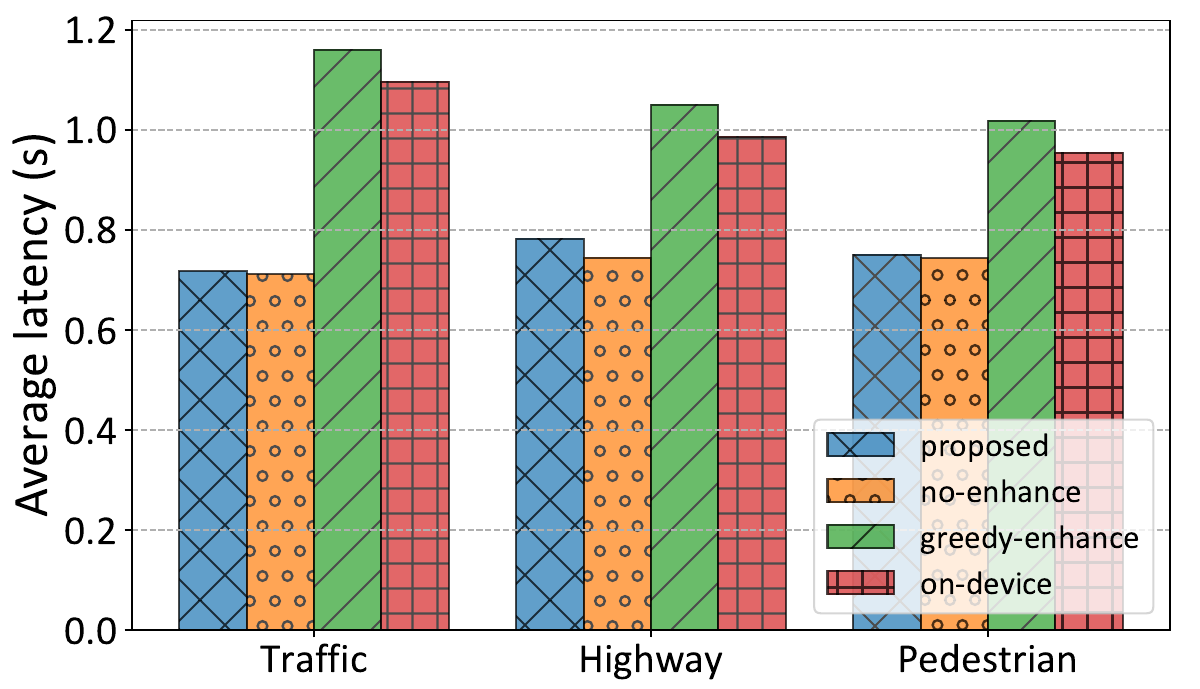}
			\label{latency_method}
			\vspace{-5mm}
		\end{minipage}
	}
	\subfigure[Average datasize of a segment.]
	{
		 \begin{minipage}[b]{.3\linewidth}
			\centering
			\includegraphics[scale=0.29]{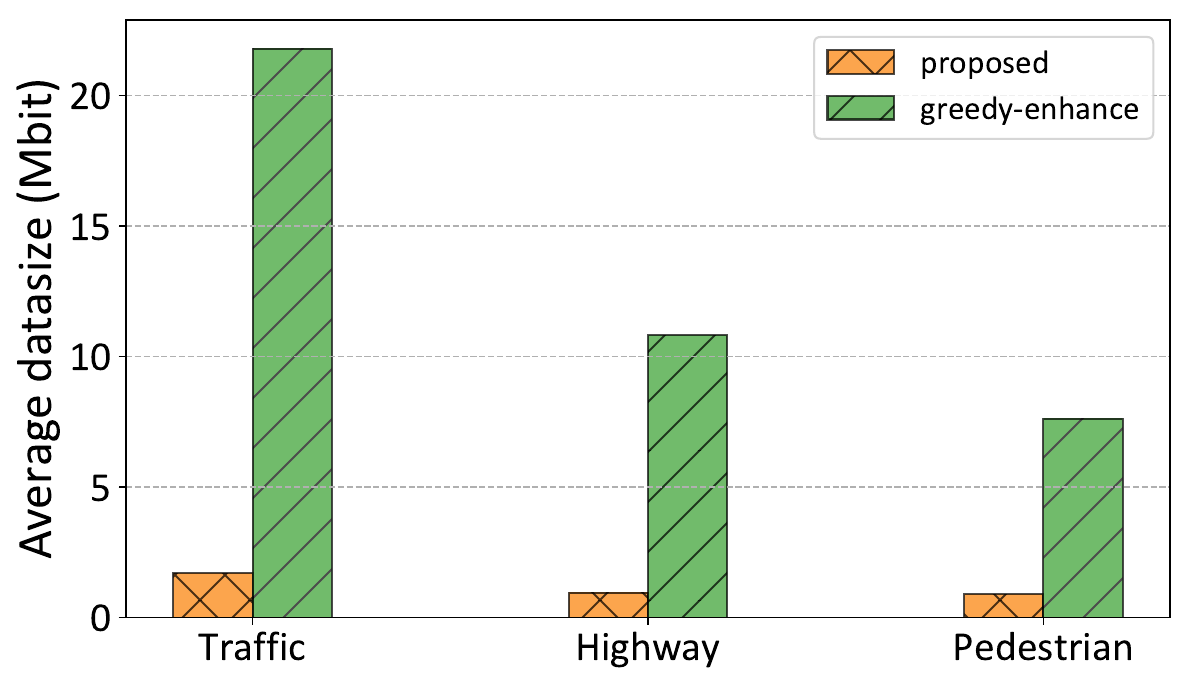}
			\label{datasize_method}
			\vspace{-5mm}
		\end{minipage}
	}
	\caption{The performance comparison of different methods on three videos.}
	\vspace{-5mm}
\end{figure*}

\begin{figure*}[ht]
	\begin{minipage}[t]{0.32\linewidth}
		\centering
		\includegraphics[width=2.2in]{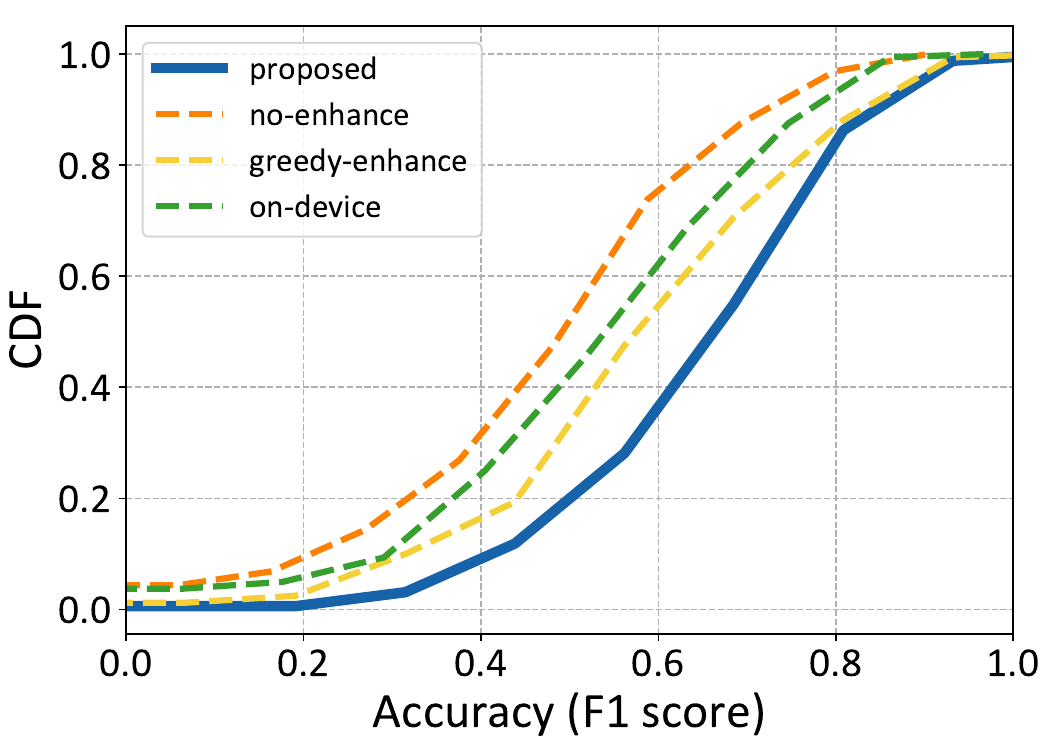}
		\vspace{-4mm}
		\caption{Accuracy of the ``Highway'' video.}
		\label{fig_cdf_accu}
	\end{minipage}
	\begin{minipage}[t]{0.36\linewidth}
		\centering
		\includegraphics[width=2.2in]{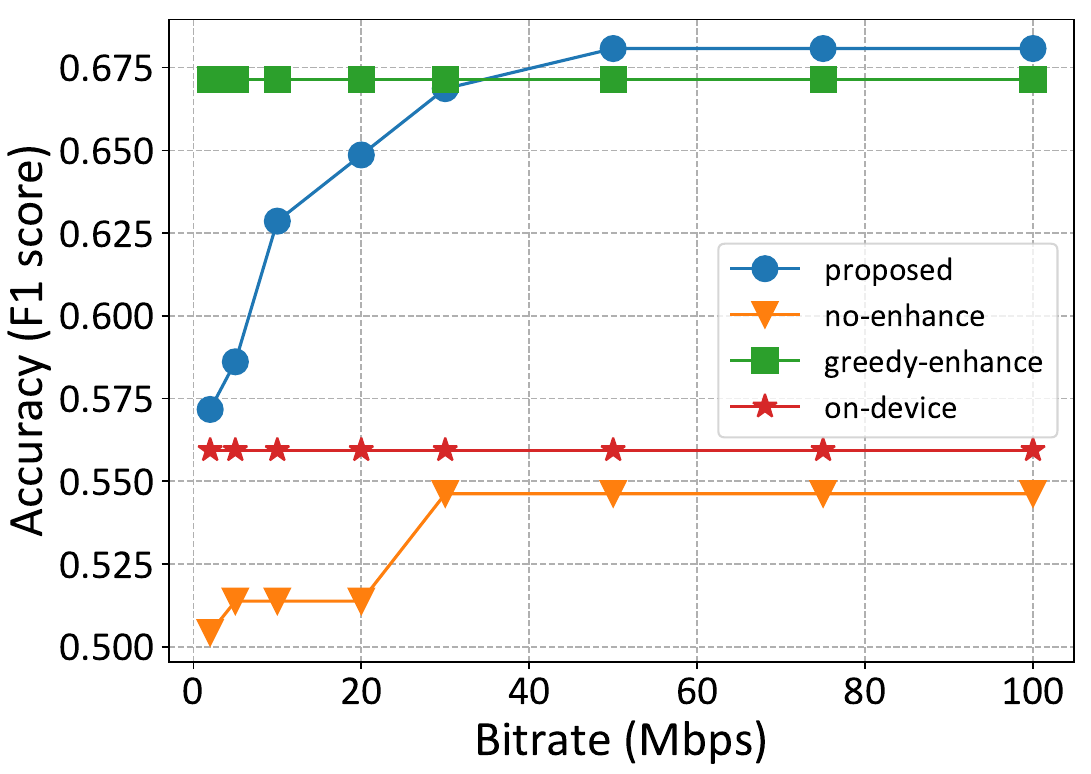}
		\vspace{-4mm}
		\caption{Impact of bandwidth condition on accuracy.}
		\label{bitrate_method}
	\end{minipage}
	\begin{minipage}[t]{0.32\linewidth}
		\centering
		\includegraphics[width=2.22in]{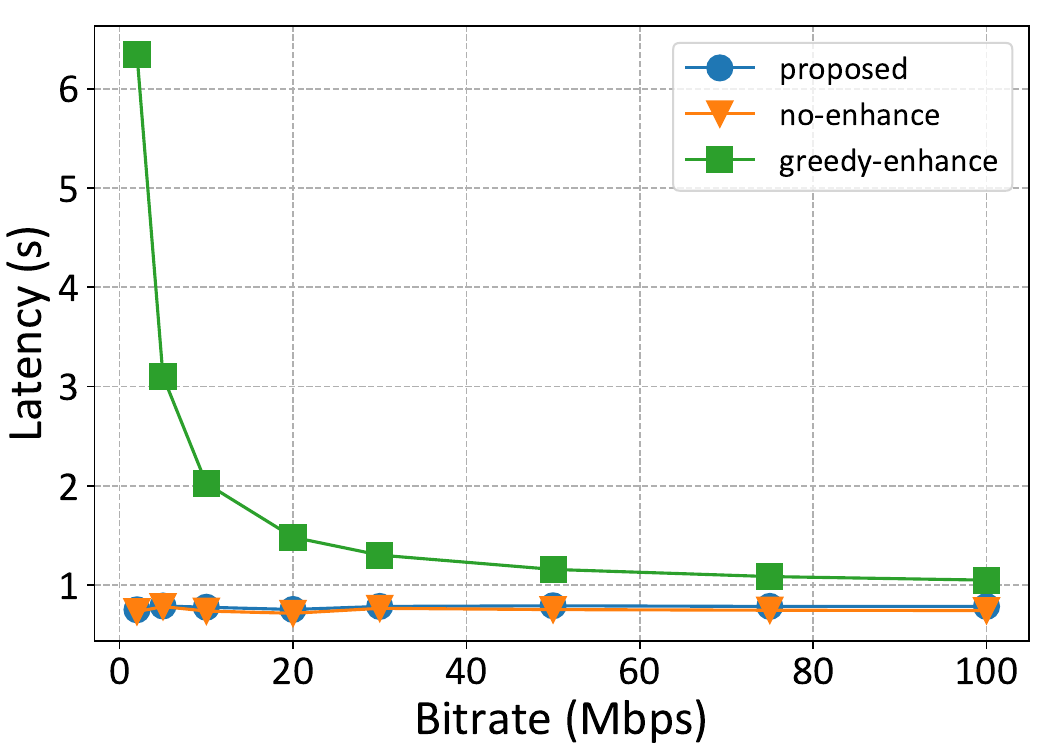}
		\vspace{-4mm}
		\caption{Impact of bandwidth condition on latency.}
		\label{bitrate_latency_method}
	\end{minipage}
\vspace{-5mm}
\end{figure*}

\subsection{Experimental Settings}
We implement the proposed system on three low-light videos, 
which are derived from YouTube\footnote{\begin{tabular}[t]{@{}l@{}}
	https://www.youtube.com/watch?v=leskMPI2dN4. Accessed May 18, 2023.\\
	https://www.youtube.com/watch?v=R7xKbFcLoXI. Accessed May 18, 2023.
\end{tabular}}. 
The resolution and frame rate of them are all 1920$\times$1080 and 
30 fps with the duration of 10 seconds. And the duration of a segment is 1 second. We called the three of them ``Traffic'', ``Highway'' and ``Pedestrian'', respectively. 
The hyper-parameters for the utility function $\{B_t,C,f,L_0,\lambda\} $ are set as \{100 Mbps, 34.1 TFLOPS, 30, 0.8 s, 0.25\} by default. 
The QP set is \{7, 12, 17, 22, 27, 32, 37, 42, 47\} corresponding to nine different quality levels and the model scale 
ratio of enhancement network are \{0, 0.25, 0.5, 0.75, 1\} corresponding to five different enhancement levels. 
The low-light videos are encoded with FFmpeg\footnote{https://ffmpeg.org/. Accessed May 18, 2023.}
on a CPU of Intel(R) Core(TM) i9-11900K @ 3.50 GHz. The object detection model for low-light video is YOLOv5s trained on COCO\cite{lin2014microsoft} dataset on NVIDIA GeForce RTX 3080 Ti GPU as edge. We compare the proposed an end-edge coordinated joint encoding and enhancement system with following baselines.
\begin{itemize}
	\item \textbf{No-enhance:} This approach adaptively streams the low-light video with different QP values according to bandwidth but does not enhance the decoded frames. 
	\item \textbf{Greedy-enhance:} This approach streams the low-light video with the highest quality level and highest enhancement level.
	\item \textbf{On-device:} It does not transmit the captured video and only performs video analysis tasks on device.
\end{itemize}

\begin{comment}
\begin{figure}[t]
	\centerline{\includegraphics[scale=0.3]{cdf_accu.pdf}}
	\caption{The cdf of accuracy of the "Highway" video.}
	\label{fig_cdf_accu}
	\vspace{-5mm}
\end{figure}

\begin{figure}[ht]
	\centering
	\subfigure[Impact of bandwidth condition on accuracy.]
	{
		\begin{minipage}[b]{.45\linewidth}
			\centering
			\includegraphics[scale=0.2]{bitrate-method.pdf}
			\label{bitrate_method}
		\end{minipage}
	}
	\subfigure[Impact of bandwidth condition on latency.]
	{
		 \begin{minipage}[b]{.45\linewidth}
			\centering
			\includegraphics[scale=0.24]{bitrate_latency_method.pdf}
			\label{bitrate_latency_method}
		\end{minipage}
	}
	\caption{The performance comparison of various bandwidth condition.}
	\vspace{-5mm}
\end{figure}

\end{comment}
\begin{table}[t]
	\label{tab:example}
	\caption{Average inference accuracy comparisons}
	\centering
	\begin{tabular}{cc}
		\toprule
		Network & Accuracy (F1 score)\\
		\midrule
		ZeroDCE\cite{guo2020zero} & \textbf{0.675}\\
		EnlightenGAN\cite{jiang2021enlightengan} & 0.585 \\
		ToDayGAN\cite{8794387} & 0.474\\
		\bottomrule
	\end{tabular}
	\vspace{-7mm}
\end{table}
\subsection{Comparison of Accuracy}
We first utilize three state-of-the-art unsupervised neural enhancement networks in our proposed system, Zero-DCE\cite{guo2020zero}, EnlightenGAN\cite{jiang2021enlightengan}, and ToDayGAN\cite{8794387}.
Table $\mathrm{I}$ shows the average inference accuracy of different methods on three video datasets. Zero-DCE achieves the best accuracy. Meanwhile, the structure of DCE-net in Zero-DCE provides opportunities to adjust model scale ratio. Therefore, we consider enhancement with Zero-DCE. We then compare the average inference accuracy of our proposed system with other three baselines. As shown in Fig. \ref{accu_method}, 
our proposed system achieves higher accuracy. To be specific, the enhancement of low-light frames improves 
the accuracy from 3.17\% to 24.63\% compared to ``No-enhance'' and ``On-device'', especially in highway scenarios. 
Compared with the other two scenes, the video content changes more frequently because of the fast movements of objects in this scene. 
In addition, compared with ``Greedy-enhance'', 
our method improves the accuracy from 1.67\% to 5.71\%. This is because ``Greedy-enhance'' cannot adjust the scale 
of the enhancement network adaptively, and the full-model is used to enhance the low-light video. We observe 
that enhancement level and quality level have dependent impact on accuracy, higher enhancement levels do not 
necessarily result in accuracy improvements, due to excessive enhancement resulting in color distortion and 
loss of details in video frames. Apart from that, the CDF of the accuracy of the ``Highway'' video is 
shown in Fig. \ref{fig_cdf_accu}. Obviously, the accuracy of our proposed method is significantly higher than the other 
three baselines. It also reveals that low-light enhancement brings much higher accuracy improvements in these scenarios.

\subsection{Comparison of Latency and Datasize}
We then compare the average latency of our proposed system with other three baselines. As shown in Fig. \ref{latency_method}, 
our proposed system achieves less latency. 
%Moreover, our proposed method achieves the latency requirement and has better accuracy improvements. 
Specifically, even though our proposed method brings a slightly longer latency than ``No-enhance'', from 
0.84\% to 5.15\%, which is due to the enhancement of low-light frames, it brings more improved accuracy than ``No-enhance''. Owing
to large datasize to transmit and insufficient computation sources, ``Greedy-enhance'' and ``On-device'' both fail to meet the 
latency requirement and they fail to adapt the quality level and the enhancement level according to bandwidth and computation constraint. 
What's more, compared with ``Greedy-enhance'', as shown in the Fig. \ref{datasize_method} the proposed method can significantly reduce the transmission datasize of the bitstream file of a segment by more than 80\%, which demonstrates the effectiveness of adaptive encoding for low-light videos.

\subsection{Impact of Bandwidth}
Next we explore the impact of bandwidth condition on the proposed algorithm in order to illustrate the effectiveness of adaptive 
encoding and enhancement. We explore different bandwidth conditions, with the maximum
available transmission bitrate from 2 Mbps to 100 Mbps. As shown in Fig. \ref{bitrate_method}, the inference accuracy of our proposed method 
improves as the transmission bitrate grows and remains stable when the bandwidth is larger than 50 Mbps. What's more, comparing to ``No-enhance'', 
our proposed method still brings much more accuracy improvements by 13.37\% when the transmission bitrate is only 2 Mbps. Even if ``No-enhance'' is stable after the transmission bandwidth is 30 Mbps, the accuracy of the proposed method 
is still improved, which also shows the effectiveness of the enhancement and makes full use of computation resources at the edge to further improve the accuracy. ``Greedy-enhance'' and ``On-device'' do not take transmission constraint into consideration, resulting in constant accuracy. In addition, the latency of ``Greedy-enhance'' reaches 6.34 s as shown in Fig. \ref{bitrate_latency_method}, 
when the transmission bitrate is 2 Mbps, which brings great obstacles to the video analysis task in low-light situation. On the contrary, our proposed 
method consistently meets the latency requirement regardless of the availability or scarcity of transmission resources, 
\begin{comment}
	because it adaptively adjusts the quality level and enhancement level to different bandwidth conditions, while utilizing the edge computation resources to enhance the low-light videos.
\end{comment}

\section{Conclusion}
In this paper, we have proposed an end-edge coordinated joint encoding and enhancement video analytics system for low-light environments. Two experiments on the impact 
on the inference accuracy as well as the communication and computation overhead to different quality levels and enhancement levels have been performed. 
A novel low-light video analytics system has been proposed to maximize the utility 
of the inference accuracy under latency constraints. And a configuration selection algorithm has been designed to adaptively stream and enhance low-light videos. Simulation experiments have shown that our proposed system achieves accuracy improvements comparing to 
no enhancement and satisfies the latency requirements, thus achieving the trade-off between communication and computation resources. For the future work, we will explore the semantic video features 
for video analytics and investigate the visual tasks in adverse environments.
\bibliographystyle{IEEEtran}	
\bibliography{ref}

\begin{thebibliography}{10}
\providecommand{\url}[1]{#1}
\csname url@samestyle\endcsname
\providecommand{\newblock}{\relax}
\providecommand{\bibinfo}[2]{#2}
\providecommand{\BIBentrySTDinterwordspacing}{\spaceskip=0pt\relax}
\providecommand{\BIBentryALTinterwordstretchfactor}{4}
\providecommand{\BIBentryALTinterwordspacing}{\spaceskip=\fontdimen2\font plus
\BIBentryALTinterwordstretchfactor\fontdimen3\font minus
  \fontdimen4\font\relax}
\providecommand{\BIBforeignlanguage}[2]{{%
\expandafter\ifx\csname l@#1\endcsname\relax
\typeout{** WARNING: IEEEtran.bst: No hyphenation pattern has been}%
\typeout{** loaded for the language `#1'. Using the pattern for}%
\typeout{** the default language instead.}%
\else
\language=\csname l@#1\endcsname
\fi
#2}}
\providecommand{\BIBdecl}{\relax}
\BIBdecl

\bibitem{olatunji2019video}
I.~E. Olatunji and C.-H. Cheng, ``Video analytics for visual surveillance and
  applications: An overview and survey,'' \emph{Neural. Comput. Appl.}, pp.
  475--515, 2019.

\bibitem{jiawei}
P.~Yang, J.~Hou, L.~Yu, W.~Chen, and Y.~Wu, ``Edge-coordinated energy-efficient
  video analytics for digital twin in \uppercase{6G},'' \emph{China
  Communications}, vol.~20, no.~2, pp. 14--25, 2023.

\bibitem{du2020server}
K.~Du, A.~Pervaiz \emph{et~al.}, ``Server-driven video streaming for deep
  learning inference,'' in \emph{Proc. of ACM SIGCOMM}, 2020, pp. 557--570.

\bibitem{yangpeng}
P.~Yang, F.~Lyu, W.~Wu, N.~Zhang, L.~Yu, and X.~Shen, ``Edge coordinated query
  configuration for low-latency and accurate video analytics,'' \emph{IEEE
  Trans. Ind. Informat.}, vol.~16, no.~7, pp. 4855--4864, 2020.

\bibitem{chengyan}
Y.~Cheng, P.~Yang, N.~Zhang, and J.~Hou, ``Edge-assisted lightweight
  region-of-interest extraction and transmission for vehicle perception,'' in
  \emph{Proc. of IEEE GLOBECOM}, 2023.

\bibitem{ananthanarayanan2019video}
G.~Ananthanarayanan, V.~Bahl \emph{et~al.}, ``Video analytics-killer app for
  edge computing,'' in \emph{Proc. of ACM MobiSys}, 2019, pp. 695--696.

\bibitem{Chuqin}
C.~Zhou, P.~Yang, Z.~Zhang, C.~Wang, and N.~Zhang, ``Bandwidth-efficient edge
  video analytics via frame partitioning and quantization optimization,'' in
  \emph{Proc. of IEEE ICC}, 2023.

\bibitem{jiang2018chameleon}
J.~Jiang, G.~Ananthanarayanan, P.~Bodik, S.~Sen, and I.~Stoica, ``Chameleon:
  scalable adaptation of video analytics,'' in \emph{Proc. of ACM SIGCOMM},
  2018, pp. 253--266.

\bibitem{10.1145/3503161.3548033}
S.~Liu, T.~Wang \emph{et~al.}, ``Adamask: Enabling machine-centric video
  streaming with adaptive frame masking for dnn inference offloading,'' in
  \emph{Proc. of ACM MM}, 2022, pp. 3035--3044.

\bibitem{9155524}
C.~Wang, S.~Zhang, Y.~Chen, Z.~Qian, J.~Wu, and M.~Xiao, ``Joint configuration
  adaptation and bandwidth allocation for edge-based real-time video
  analytics,'' in \emph{Proc. of IEEE INFOCOM}, 2020, pp. 257--266.

\bibitem{Kong}
Y.~Kong, P.~Yang, and Y.~Cheng, ``Edge-assisted on-device model update for
  video analytics in adverse environments,'' in \emph{Proc. of ACM MM}, 2023.

\bibitem{wang2022self}
W.~Wang, Z.~Xu, H.~Huang, and J.~Liu, ``Self-aligned concave curve:
  Illumination enhancement for unsupervised adaptation,'' in \emph{Proc. of ACM
  MM}, 2022, pp. 2617--2626.

\bibitem{coltuc2006exact}
D.~Coltuc, P.~Bolon, and J.-M. Chassery, ``Exact histogram specification,''
  \emph{IEEE Trans. Image Process.}, vol.~15, no.~5, pp. 1143--1152, 2006.

\bibitem{7782813}
X.~Guo, Y.~Li, and H.~Ling, ``Lime: Low-light image enhancement via
  illumination map estimation,'' \emph{IEEE Trans. Image Process.}, vol.~26,
  no.~2, pp. 982--993, 2017.

\bibitem{tao2017llcnn}
L.~Tao, C.~Zhu \emph{et~al.}, ``Llcnn: A convolutional neural network for
  low-light image enhancement,'' in \emph{Proc. of IEEE VCIP}, 2017, pp. 1--4.

\bibitem{jiang2021enlightengan}
Y.~Jiang, X.~Gong, D.~Liu, Y.~Cheng \emph{et~al.}, ``Enlightengan: Deep light
  enhancement without paired supervision,'' \emph{IEEE Trans. Image Process.},
  vol.~30, pp. 2340--2349, 2021.

\bibitem{guo2020zero}
C.~Guo, C.~Li \emph{et~al.}, ``Zero-reference deep curve estimation for
  low-light image enhancement,'' in \emph{Proc. of IEEE CVPR}, 2020, pp.
  1780--1789.

\bibitem{richardson2011h}
I.~E. Richardson, \emph{The H. 264 advanced video compression standard}.\hskip
  1em plus 0.5em minus 0.4em\relax John Wiley \& Sons, 2011.

\bibitem{chen2018learning}
C.~Chen, Q.~Chen, J.~Xu, and V.~Koltun, ``Learning to see in the dark,'' in
  \emph{Proc. of IEEE CVPR}, 2018, pp. 3291--3300.

\bibitem{cai2018learning}
J.~Cai, S.~Gu, and L.~Zhang, ``Learning a deep single image contrast enhancer
  from multi-exposure images,'' \emph{IEEE Trans. Image Process.}, vol.~27,
  no.~4, pp. 2049--2062, 2018.

\bibitem{li2023fast}
P.~Li, G.~Cheng \emph{et~al.}, ``Fast: Fidelity-adjustable semantic
  transmission over heterogeneous wireless networks,'' in \emph{Proc. of IEEE
  ICC}, 2023.

\bibitem{10000829}
C.~Wang, P.~Yang, J.~Lin, W.~Wu, and N.~Zhang, ``Object-based resolution
  selection for efficient edge-assisted multi-task video analytics,'' in
  \emph{Proc. of IEEE GLOBECOM}, 2022, pp. 5081--5086.

\bibitem{metropolis1953equation}
N.~Metropolis, A.~W. Rosenbluth \emph{et~al.}, ``Equation of state calculations
  by fast computing machines,'' \emph{The Journal of Chemical Physics},
  vol.~21, no.~6, pp. 1087--1092, 1953.

\bibitem{lin2014microsoft}
T.~Y. Lin, M.~Maire, S.~Belongie, J.~Hays \emph{et~al.}, ``Microsoft coco:
  Common objects in context,'' in \emph{Proc. of ECCV}, 2014, pp. 740--755.

\bibitem{8794387}
A.~Anoosheh, T.~Sattler \emph{et~al.}, ``Night-to-day image translation for
  retrieval-based localization,'' in \emph{Proc. of ICRA}, 2019, pp.
  5958--5964.

\end{thebibliography}
%\end{thebibliography}

\end{document}